\documentclass[draft]{agujournal}
\usepackage{amssymb,amsmath}

\begin{document}

%
%


\title{Results from The Latin American Giant
Observatory Space Weather Simulation Chain}

%
%




\authors{H. Asorey\affil{1,2}, L. A. N\'u\~nez\affil{3,4}, 
M. Su\'arez-Dur\'an\affil{3}}


\affiliation{1}{Laboratorio Detecci\'on de Part\'{\i}culas y Radiaci\'on (CNEA,CONICET,UNCUYO),  San Carlos de Bariloche, Argentina.}

\affiliation{2}{Instituto de Tecnolog\'{\i}as en Detecci\'on y Astropart\'{\i}culas (CNEA,CONICET,UNSAM), Buenos Aires, Argentina.}

\affiliation{3}{Escuela de F\'isica, Universidad Industrial de 
Santander, 680002 Bucaramanga, Colombia.}

\affiliation{4}{Departamento de  F\'isica, Universidad de Los Andes 5101 M\'erida, Venezuela.}


\correspondingauthor{M. Su\'arez-Dur\'an}
{mauricio.suarez@correo.uis.edu.co}

\begin{keypoints}
\item An extended rigidity cutoff is calculated under quiet (secular) \& transient conditions for the geomagnetic field.
\item Penumbra region is reinterpreted as a probability function of the arrival direction.
\item The effect of the geomagnetic field in the flux of Galactic Cosmic Rays can be estimated from changes on the flux of secondary particles at ground level. 
\end{keypoints}

\begin{abstract}
  The Space Weather program of the Latin American Giant Observatory
  (LAGO) Collaboration was designed to study the variation of the flux of
	atmospheric secondary particles at ground level produced during the
	interaction of cosmic rays with the air. This work complements and
	expands the inference capabilities of the LAGO detection network to
	identify the influence of solar activity on the particle flux, at
	places having different geomagnetic rigidity cut-offs and atmospheric
	depths. This program is developed through a series of Monte Carlo
	sequential simulations to compute the intensity spectrum of the various
	components of the radiation field on the ground. A key feature of these
	calculations is that we performed detailed radiation transport
	computations as a function of incident direction, time, altitude, as
	well as latitude and longitude.  Magnetic rigidity calculations and
	corrections for geomagnetic field activity are established by using the
	MAGNETOCOSMICS code, and the estimation of the flux of secondaries at
	ground level is implemented by using the CORSIKA code; thus we can
	examine the local peculiarities in the penumbral regions with a more
	realistic description of the atmospheric and geomagnetic response in
	these complex regions of the rigidity space. As an example of our
	calculation scheme, we report some result on the flux at ground
	level for two LAGO locations: Bucaramanga-Colombia and San Carlos de
	Bariloche-Argentina, for the geomagnetically active period of May
	2005.
\end{abstract}

\section{Introduction}

Solar activity has a strong influence on the modulation of the
flux of galactic cosmic rays (GCRs) arriving at Earth, whose
transport through the heliosphere is one of the topics of 
major interest in space physics, presenting several unresolved  
questions. Space weather physics is experiencing a fast growing 
interest nowadays because of evidence that environmental 
conditions in the near-Earth space have direct and indirect 
impacts on technology and the global 
economy\,\citep{Schrijver2015}.

One of most puzzling modulation of the cosmic rays flux at 
ground level is called Forbush Decrease (FD): a rapid 
reduction in the observed galactic cosmic ray intensity followed by 
a slow exponential-like recovery\,\citep{Usoskin2008}. This phenomenon 
initially was reported by S.E. Forbush --and almost simultaneously by 
V.F. Hess \& A. Demmelmair-- in 1937\,
\citep[see e.g.][]{Forbush1937, victor-demmelamair1937}. Later, in the
1950's, the works of Simpson, Fonger \&
Treiman\,\citep{SimpsonFongerTreiman1953}, and Singer\,\citep{Singer1954},
showed that FDs are related to the solar activity interacting with the
interplanetary medium. More recently, Lockwood showed a dependence of the
magnitude of the FD upon the vertical cutoff
rigidity\,\citep{Lockwood1971}.

FDs can be classified in two groups: recurrent and non-recurrent FD.  While the
first group\,\citep{Lockwood1971} have a symmetric profile and are well
associated with co-rotating high speed solar wind streams\,\citep{Cane2000},
non-recurrents FD have sudden onsets with a maximum depression within a day,
and a more gradual recovery\,\citep{Cane2000, Belov-etal2014}
, and are related to the interaction of GCRs, Interplanetary Coronal Mass Ejections
(ICME) and a perturbed geomagnetic field (it is perturbed by its
interaction with the same ICME). 

Understanding these very complex phenomena depends upon: in situ measurements
of the interaction of GCRs-ICME; tracking the propagation of GCRs through the
Geomagnetic Field (GF hereafter); taking into account their
interaction with the atmosphere, and also the variation of
the particles produced by the latter interaction (hereafter secondaries) 
at ground level.

As the secondaries are produced by the interaction of GCRs with 
the atmosphere (hereafter denoted as primaries), the modulation of
secondary particles needs to be monitored and carefully corrected
by taking into account several atmospheric factors that could 
modify the flux of secondaries at Earth's
surface (the Atmospheric density profile is one
of those factors because this is proportional to the absorption
of secondaries
particles)\,\citep{ThePierreAugerCollaboration2011,
2012AdSpR..49.1563D, Asorey2013, Auger2015,Suarez2015}. Furthermore,
a series of completed and detailed simulations are 
needed to characterize the expected flux at the detector
altitude. This kind of simulation must take into account several
important factors such as GF conditions, i.e., estimations of the
magnetic rigidity of the GCRs, the interaction of GCRs with
atmosphere, the variations in atmospheric depth and the
detector response. 

Direct solar wind observations using spacecraft can provide 
insight of the interplanetary magnetic field; but its 
global structure can not be completely monitored by on-board 
measurements, because they can only detect local variations 
along the trajectory of the probe within the solar wind. On 
the other hand, ground observatories with detectors spread 
on very large areas, registering indirectly low energy GCRs, 
can provide important, alternative and complementary 
information about the broader structure of the interplanetary 
magnetic fields and its influence on the GCRs flux. In this 
context, a global network of both type of observatories combines
different techniques to monitor the development of FDs at
different geomagnetic latitudes and rigidity cutoffs, will enrich
future studies on the detection of solar modulation of
GCRs\,\citep[see e.g.][]{Abbasi2008,Auger2015}.

The Latin American Giant Observatory 
(LAGO)\,\citep{Allard-etal2008, Asorey2015a} has developed a 
program to understand the influence of the space weather 
phenomena on the flux of GCRs. This program, called LAGO Space 
Weather (LAGO-SW)\,\citep{Asorey2013}, includes a precise
simulation that takes into account the geomagnetic corrections
and a detailed
measurement of the modulation on the flux of secondaries, and 
evaluates if this modulation have possible causal 
correlations with space weather phenomena, like 
FDs\,\citep{Suarez2015}.

Nowadays, computational capabilities allow the extension of 
the usual approach, which is to consider only the components
of the GCR flux locally and include geomagnetic effects by an 
effective rigidity cutoff for vertical
primaries\,\citep{MasiasDasso2014}. The detailed
simulations described in this work, include these effects over
different arrival directions during dynamic events affecting the
geomagnetic field and
atmospheric conditions. We have generalized previous attempts by 
including not only secular, but also transient variations of 
the directional geomagnetic rigidity cutoff.

This paper is organized as follows: in section \ref{LAGO}, 
the Latin American Giant Observatory and its space weather 
program are briefly described; then, in Section \ref{chain}, we 
present our space weather simulation chain, focusing on 
geomagnetic corrections of the primary flux; in Section 
\ref{results} we discuss our main results in secular conditions 
and later under geomagnetic disturbances for two LAGO sites: 
Bucaramanga-Colombia and San Carlos de Bariloche-Argentina. 
Finally, in section \ref{conclusions} some finals remarks and 
future projects are considered.

\section{The LAGO Space Weather Program}
\label{LAGO}
The Latin American Giant Observatory (LAGO) is an extended
astroparticle observatory on a continental scale, promoting
training and research in astroparticle physics in Latin America
covering three main areas: search for the high energy component of
gamma rays bursts (GRBs) at high altitude sites, space weather
phenomena, and background radiation at ground
level\,\citep{Asorey2015a}.

\begin{figure}[h]
	\centering
	\includegraphics[width=32pc]{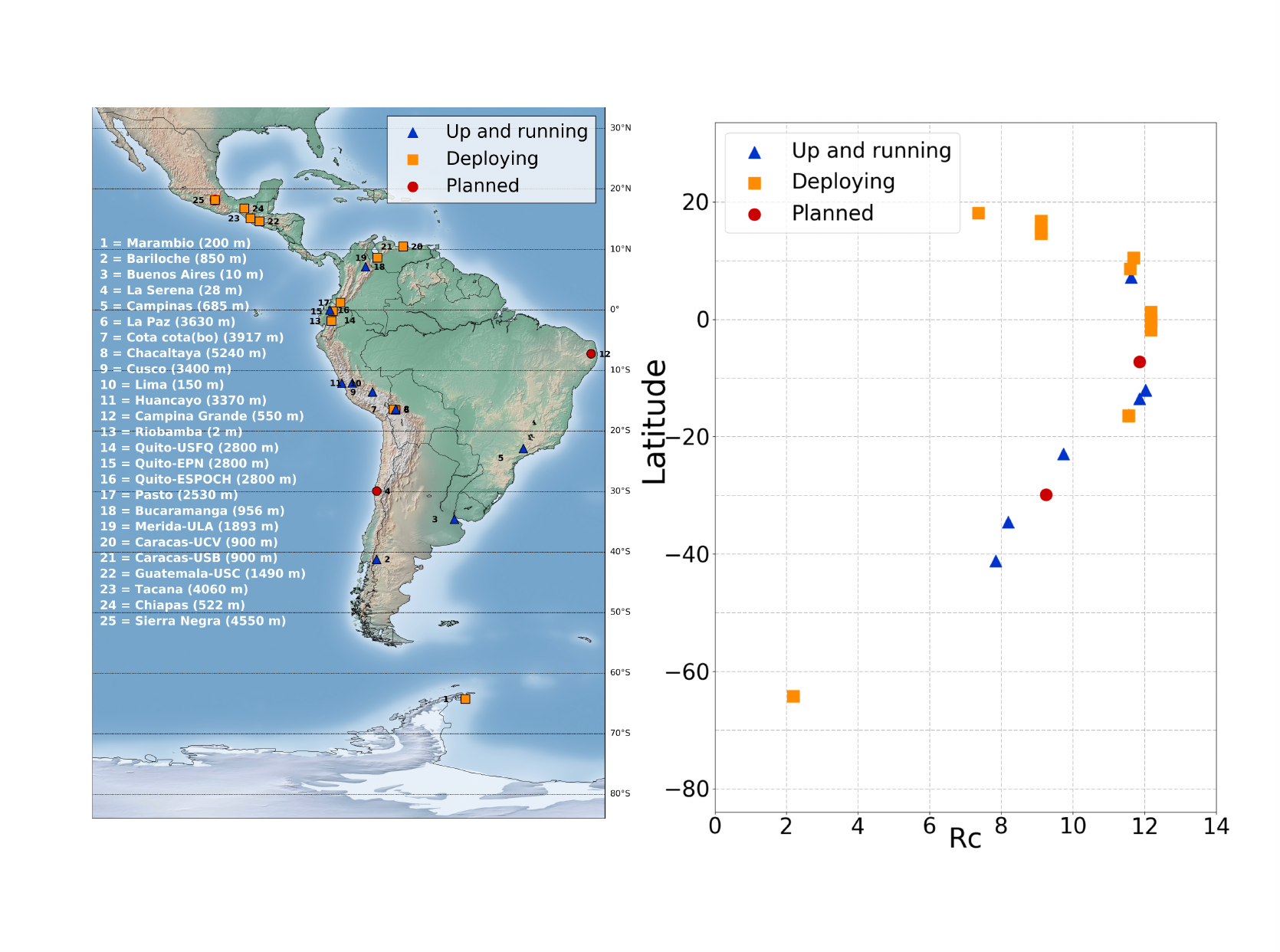}
	\caption {Left panel indicates the geographical
  distribution and altitudes of LAGO water Cherenkov detectors:
  operational are represented by blue triangles; in deployment by
  orange squares and finally, red circles indicating planned
  sites. The right panel shows the vertical rigidity cutoff at
  each LAGO site.}
	\label{lago-sites}
\end{figure}

The LAGO detection network consists of ground-level
water-Cherenkov particle detectors (WCDs), spanning over several
sites, located at significantly different latitudes and various
altitudes --from Mexico to the Patagonia and from mean sea level
up to more than 5000\, meters of altitude. After the installation
of new detectors at the Antarctica
Peninsula\,\citep{DassoEtal2015}, LAGO will cover a large range
of geomagnetic rigidity cutoffs and atmospheric
absorption/depths\,\citep{Sidelnik2015}. The current/planned
distribution and status of the LAGO detection network is shown in
Figure \ref{lago-sites}. This network of detectors is operated by
the LAGO Collaboration: a non-centralized and distributed
collaborative network of more than 80 scientists from
institutions of te Latin American countries (Argentina,
Bolivia, Brazil, Chile, Colombia, Ecuador, Guatemala, Mexico, Peru and
Venezuela) and Spain. The LAGO Collaboration is using WCDs in all
sites, due to
their proved reliability, high detection, low cost and efficiency
of the detection of all components present in atmospheric
extensive showers\,\citep{Asorey2015a, Sidelnik2015,
Galindo2015, DassoEtal2015}.

As explained in \citet{Asorey2015a}, the LAGO scientific and
academic objectives are organized in different programs. The
Space Weather LAGO Program is devised to study variations in the
flux of secondary particles at ground level and its relation to
the heliospheric modulation of GCRs. The LAGO detector network
determines the flux of secondary particles in different bands of
deposited energy in the detector, by using pulse shape
discrimination techniques. This is what we have called the
multi-spectral analysis technique\,\citep{Suarez2015}. The
total energy threshold for the detection of secondary particles
are $\simeq 0.4$\,MeV for gammas, $\simeq 0.8$\,MeV for electrons
and $\simeq 160$\,MeV for muons.

By combining all the data measured 
at different locations, LAGO provides simultaneous and
detailed information of the temporal evolution of the secondary
flux at different geomagnetic locations. This can help to get a
better understanding of the small and large spacetime scales of
the disturbances produced by different space weather
phenomena\,\citep{Suarez2015}.

Any attempt to estimate the expected flux of secondaries at any
detector of the LAGO network should be based on a detailed
simulation that takes into account all possible sources of flux
variation. This complex approach comprises processes occurring at
different spatial and time scales, following this conceptual
scheme:
\begin{center}
    \begin{tabular}{ccccc}
    GCR Flux & $\xrightarrow{\mathrm{Heliosphere}}$ &
    Modulated Flux & $\xrightarrow{\mathrm{Magnetosphere}}$ &
    Primaries $\rightarrow \cdots$ \\ $\cdots \rightarrow$ 
    Primaries & $\xrightarrow{\mathrm{Atmosphere}}$ & 
    Secondaries & $\xrightarrow{\mathrm{Detector}}$ & 
    Signals.
    \end{tabular}
\end{center}
As it can be easily appreciated, the above simulation pipeline
considers three important factors with different spatial and time
scales: the geomagnetic effects, the development of the extensive
air showers in the atmosphere, and the detector response at
ground level. According to the above scheme, in this work we
focused in
stages covering from the modulated flux to the flux of secondary
particles at
ground level.

Our {\it Simulation Chain} can be depicted in three main consecutive blocks:
\begin{enumerate}
	\item The effects of GF on the propagation of charged
    particles, contributing to the background radiation at ground
    level, are characterized by the directional rigidity cutoff,
    $R_\mathrm{C}$, at each LAGO site and calculated using the
    MAGNETOCOSMICS code\,\citep{Desorgher2004} applying the
    backtracking technique\,\citep[see e.g.][]{MasiasDasso2014}.
    The Geomagnetic Field at any point on Earth is determined by
    using the International Geomagnetic Field Reference, version
    11\,\citep{IGRF11} at the near-Earth GF ($r<5
    R_{\oplus}$) --$r$ distance from Earth center and
    $R_{\oplus}$ is the Earth radius ($6371$\,km)-- and through
    the Tsyganenko Magnetic Field model version 2001 (TSY01
    hereafter)\,\citep{TSY01a} to describe the outer GF ($r > 5
    R_{\oplus}$).

	\item The second step of the chain is based on the CORSIKA
    code\,\citep{Corsika}. Extensive air showers produced during
    the interaction of cosmic rays with the atmosphere are
    simulated with extreme detail to obtain a very comprehensive
    set of secondaries at ground level.
  
\item Finally, a GEANT4 model\,\citep{Agostinelli2003} of the detector
	response to the different types of secondary particles is being
		implemented\,\citep{Otiniano2015, Vargas2015, CalderonAsoreyNunez2015}
		and will be reported in the near future.
\end{enumerate}

\section{The space weather simulation chain}
\label{chain}
The propagation of charged particle through the GF has been
studied since
the 60s and was focused on understanding how the penumbra region changes with
the geographical position\,\citep[see e.g.][]{shea-smart-mccracken-1965,
Carmicheal-etal1969, smart-shea2012}. In this section we shall discuss our
novel approach to understand the penumbra region and our proposal for a new
method to calculate the magnetic rigidity as a function of time. We shall also
describe in detail how geomagnetic effects on the low energy flux of primaries
can be infered from observations of secondary particles at ground level by
means of the following procedure:
\begin{enumerate}
  \item To find a magnetic rigidity function, $R_\mathrm{m}(\mathrm{Lat},
	\mathrm{Lon}, \mathrm{Alt}, t, \theta, \phi)$, at a particular
		geographical position --i.e. latitude (Lat), longitude (Lon) and
		altitude above sea level (Alt)--, time ($t$) and arrival direction of
		the particle --i.e. zenithal ($\theta$) and azimuthal ($\phi$) angle;
	\item To calculate the flux of primaries at the top of the
    atmosphere ($\approx 112$\,km a.s.l. (Above
    Sea Level)), filtered by the magnetic rigidity function
		$R_\mathrm{m}(\mathrm{Lat},
    \mathrm{Lon}, \mathrm{Alt}, t, \theta, \phi)$;
	\item To estimate the flux of secondaries at ground level produced by
		the interactions of the impinging GCRs with the atmosphere.
\end{enumerate}

The following subsections will develop all details for the above-mentioned actions.

\subsection{Magnetic rigidity as function of time}
\label{sub:rm-time}

The direction of the velocity of a GCR ($\hat{I}=\vec{v}/|\vec{v}|$) changes
along the particle trajectory inside the dynamic GF,
$\mathbf{\vec{B}(\mathrm{Lat}, \mathrm{Lon}, \mathrm{Alt}, t)}$,
according to the equation
\begin{linenomath*}
	\begin{equation}\label{def:rm}
    \frac{d\hat{I}}{ds} = \frac{1}{R_\mathrm{m}}
    \left( \hat{I} \times \mathbf{\vec{B}(\mathrm{Lat}, \mathrm{Lon},
    \mathrm{Alt}, t)} \right)\, ,
	\end{equation}
\end{linenomath*}
where $s$ is the path length along the particle trajectory and
$R_\mathrm{m}=pc/Ze$ the magnetic rigidity; with $p$ as the particle
momentum, $c$ the light speed, $Z$ the atomic number and $e$ the
electric charge of the electron. The variation of $\hat{I}$ is
weighted by $R_\mathrm{m}$ and therefore a GCR is able to arrive
at some specific geographical point --under some configuration of
$\mathbf{\vec{B}(\mathrm{Lat}, \mathrm{Lon}, \mathrm{Alt}, t)}$
associated with the trajectory of the particle arriving to a
particular position, i.e Latitude ($\mathrm{Lat}$), Longitude
($\mathrm{Lon}$), Altitude ($\mathrm{Alt}$)-- if its
$R_\mathrm{m}$ has the right value. Thus, we can write
$R_\mathrm{m}$ as
\begin{linenomath*}
	\begin{equation}
		\label{def:newrm}
    R_\mathrm{m} \equiv R_\mathrm{m}(\mathrm{Lat}, \mathrm{Lon}, 
        \mathrm{Alt},t)\,.
    \end{equation}
\end{linenomath*} 

Following standard definitions\,\citep[see e.g.][]
{Cooke-etal1991}, particles with \textit{allowed} $R_\mathrm{m}$
will reach at certain geographical position, while those with
\textit{forbidden} $R_\mathrm{m}$ will not. With these
considerations, three different ranges of $R_\mathrm{m}$ can be
defined:
\begin{itemize}
	\item \textit{Forbidden} range: a continuous range which goes from zero to
		the first \textit{allowed} value of $R_\mathrm{m}$, say $R_\mathrm{L}$;
	\item \textit{Allowed} range: cotaining all the rigidities above a certain
		$R_\mathrm{m}$ value, say $R_\mathrm{U}$, for which all the rigidities
		containted in this range are \textit{allowed}.
	\item \textit{Penumbra} range: the range ($R_\mathrm{L} < R_\mathrm{m}
		< R_\mathrm{U}$) connecting the \textit{allowed} and \textit{forbidden}
		ranges.
\end{itemize}

The \textit{penumbra} is characterized by a single, effective,
rigidity value\,\citep{shea-smart-mccracken-1965, SmartShea2009},
which is used to establish whether a GCR arrives, or not, at the
particular geographical point. This value is called \textit{the
rigidity cutoff} ($R_\mathrm{C}$) and can be defined as
\begin{linenomath*}
	\begin{equation}
		\label{def:basic-rc}
    R_\mathrm{C} = R_\mathrm{U} -
    \sum_{k=R_\mathrm{L}}^{R_\mathrm{U}} \Delta
    {R_k}^{\mathrm{allowed}}\, ,
	\end{equation}
\end{linenomath*}
where $\Delta{R_k}$ is the resolution of the $R_\mathrm{m}$
calculation. Strictly speaking, $R_\mathrm{U}$ and $R_\mathrm{L}$
depend on time, the arrival direction, the geographical position
and the altitude; thus, we should consider that, at a
geographical point,
 
\begin{linenomath*}
	\begin{equation}
		\label{def:renewrm}
    R_\mathrm{m} = R_\mathrm{m}\left(
		\mathrm{Lat},\mathrm{Lon},\mathrm{Alt},t,\theta,\phi
		\right)\, ,
	\end{equation}
\end{linenomath*}

It is important to note that definition (\ref{def:basic-rc})
has the implicit assumption that all the particle trajectories
have the same contribution in the penumbra region, i.e., the
flat GCR spectrum approximation, according
to\,\citep{Dorman-etal2008}. In this approximation, the very
complex problem of allowed trajectories in the penumbra region is
simply  replaced by an effective cutoff, only calculated for
vertical primaries. We have refined this approximation by
considering the penumbra not as a sharp cutoff, but as a
relatively smooth transition between the forbidden and the
allowed regions. In our approach, we extend the concept of the
effective rigidity cutoff assuming that it can be approximated by
a cumulative probability function (CDF).

The next subsection outlines the method we have implemented to
calculate $R_\mathrm{m}$ and to characterize the penumbra region
as a cumulative probability function (CPF).

\subsubsection{Magnetic Rigidity Calculation}
\label{subsubsec:rm-calcu}

We performed the $R_\mathrm{m}$ calculation by the backtracking
technique\,\citep[see e.g.][]{MasiasDasso2014}, via the
MAGNETCOSMICS (MAGCOS) code, with a resolution of $0.01$\,GV,
considering two conditions: secular and dynamic geomagnetic field
effects. For secular conditions we used the configuration of
the geomagnetic field on $6$\,UTC April 26$^{th}$ 2005, because at
this time the registered Disturbance Storm Time Index (Dst index
hereafter,\\\noindent https://www.ngdc.noaa.gov/stp/geomag/dst.html)
was zero with a variability of $0.79$\,nT from 0 UTC of April
26$^{th}$ to 12 UTC of the same day 
(http://wdc.kugi.kyoto-u.ac.jp/dst\_final/200504/index.html).
For the dynamic GF contribution, we calculated the $R_\mathrm{m}$
according to the GF configuration for each hour of May 2005,
setting the parameters: Dynamic pressure, Dst index,
interplanetary magnetic fields components $B_x$ and $B_z$, and
Tsyganenko's parameters for model TSY01: G1 and G2\citep{TSY01a}.

These parameters were taken from the Virtual Radiation Belt
Observatory\,\citep{Weigel-etal2009}; $R_\mathrm{m}$ values were
calculated for zenith angles from $0^\circ$ to $90^\circ$, with
$\Delta\theta = 15^\circ$ and azimuth angles (for each $\theta$)
from $0^\circ$ to $360^\circ$ with $\Delta\phi=15^\circ$,
for both site.

\subsubsection{Interpreting the Penumbra Region}
\label{subsubsec:inter-penum}
Instead of the standard simplifying assumption for the 
penumbra region we build a cumulative probability function 
(CPF), valid from $R_\mathrm{L}$ to $R_\mathrm{U}$, which
replaces the usual concept of $R_\mathrm{C}$, (defined in
equation (\ref{def:basic-rc})). We denote this CPF as
$P(R_\mathrm{m}(\theta))$, which represents the probabilty of the
cosmic ray arriving at
some geographical position with zenith angle $\theta$, at time
$t$, with $R_\mathrm{m}$;
we take into account the following considerations: 

\begin{itemize}
	\item the backtracking technique performed by MAGCOS is 
    a deterministic method, which implies that it is not 
    possible to calculate a statistical set of $R_\mathrm{m}$
    values for a specific arrival direction, i.e., pair of
    ($\theta$,$\phi$);
	\item for each zenith angle we consider $24$ uniform ranges in azimuth with
		an angular amplitude of $15^{\mathrm{o}}$ each one, i.e. for each zenith
		angle we have $24$ different penumbra regions; and 
	\item for each $\theta$, the associated set of $R_\mathrm{m}$ in the $24$
		penumbra regions has a global minimal value ($R_\mathrm{Lmin}$
		hereafter) and a global maximum value ($R_\mathrm{Umax}$ hereafter).
\end{itemize}

Accordingly, it is possible to come up with a frequentist approach,
assuming a probability function defined as:
\begin{linenomath*}
	\begin{equation}
		\label{def:pdf}
    \wp(R_\mathrm{m}(\theta)) =
    \frac{\#R_\mathrm{m_{\mathrm{allowed}}}(\theta)}
    {\#R_\mathrm{m_{\mathrm{tot-allowed}}}(\theta)}\, ,
	\end{equation}
\end{linenomath*}
where, for each $\theta$, we have averaged over the azimuth angle, 
$\phi$, within each penumbral range; the fraction of
the number ($\#$) of allowed $R_\mathrm{m}$ values
($R_\mathrm{m_{allowed}}(\theta)$) over the total number of $R_m$
values calculated for the $\theta$'s set
($R_\mathrm{m_{tot-allowed}} (\theta)$).

Equation (\ref{def:pdf}) implies that the domain interval for
$R_\mathrm{m_{tot-allowed}}$ is
\begin{linenomath*}
	\begin{equation}
		\label{def:domain-cdf}
    \mathcal{D}(R_{\mathrm{m}_{\mathrm{tot-allowed}}}) = \left\{
      R_\mathrm{m}(\theta): \left( R_\mathrm{m}\geq
      R_{\mathrm{Lmin}}, \, 
      R_\mathrm{m} \leq R_{\mathrm{Umax}} \right) \right\}\, .
    \end{equation}
\end{linenomath*}

Thus, from (\ref{def:pdf}) and (\ref{def:domain-cdf}) we 
define the cumulative distribution function for a GCR, 
arriving at the observation point with rigidity
$R_\mathrm{m}(\theta)$ as
\begin{linenomath*}
	\begin{equation}
		\label{def:cdf}
    P(R_\mathrm{m}(\theta)) = \sum_{\mathcal{R}_m =
		\mathrm{R_{Lmin}}}^{\mathcal{R}_m =
		\mathrm{R_{Umax}}}\wp(\mathcal{R}_m(\theta))\, .
	\end{equation}
\end{linenomath*}
Notice that equation (\ref{def:cdf}) implies that a GCR with 
$R_\mathrm{m} > R_\mathrm{Umax}$ has a probability of $1$ to
arrive at the observation point through the zenith angle
$\theta$; meanwhile a GCR with
$R_\mathrm{m}<R_\mathrm{Lmin}$ has $0$ probability to arrive at
the same point with the same angle.

Currently, the usual $R_\mathrm{C}$ is interpreted as a unique
value in the penumbra region, which separates only two
possibilities for an incoming particle with a zenith angle
$\theta$: arriving or not arriving. If a charged particle has a
$R_\mathrm{m}~>~R_\mathrm{C}$ then it is considered to arrive at
the geographical point, in the opposite case, if
$R_\mathrm{m}~<~R_\mathrm{C}$, it does not arrive. With our
approach, it is clear that a GCR, with $R_\mathrm{m}$ and zenith
angle $\theta$, can reach at the geographical point if
$P(R_\mathrm{m}(\theta))~=~1$, whereas with
$P(R_\mathrm{m}(\theta))~=~0$ will not. But, if the
$R_\mathrm{m}$ belongs to the penumbra region, it does not meet
any of the above criteria because $P(R_\mathrm{m}(\theta))$ is
between 0 and 1. To set this value of $P(R_\mathrm{m}(\theta))$
in terms of arriving or not arriving, i.e., 0 or 1, we implement
a Metropolis Monte Carlo algorithm as follows: for a $P(R_\mathrm{m}(\theta))$ 
value, different from $1$ and $0$, we calculate a random number: $0~<~P_\mathrm{temp}~<~1$. Then, 
\begin{itemize}
  \item if $P(R_\mathrm{m}(\theta)) \geq P_\mathrm{temp}$, then
    $P(R_\mathrm{m}(\theta))=1$; otherwise
	
  \item if $P(R_\mathrm{m}(\theta)) < P_\mathrm{temp}$, then
    $P(R_\mathrm{m}(\theta))=0$.
\end{itemize}
Therefore, we interpret the rigidity cutoff $R_\mathrm{C}$ as
function of the cumulative distribution function, i.e., 
\begin{linenomath*}
	\begin{equation}
		\label{def:rc-sec-cdf}
    R_\mathrm{C} = R_\mathrm{C}(\mathrm{Lat}, \mathrm{Lon},
    \mathrm{Alt}, \theta, P(R_\mathrm{m}(\theta)) ) \, .
	\end{equation}
\end{linenomath*}

Now, from the dynamic magnetic rigidity definition we perform the
same type of calculations but including the time ($t$)
dependence, by evaluating equation (\ref{def:renewrm}) for
different conditions at particular moments.

After applying the same procedure, we obtained the dynamic
rigidity cut-off of a site, as
\begin{linenomath*}
	\begin{equation}
		\label{def:rc-dyn}
    R_\mathrm{C} = R_\mathrm{C}(\mathrm{Lat}, \mathrm{Lon},
    \mathrm{Alt}, \theta, t, P(R_\mathrm{m}(\theta),
		t)) \, ,
	\end{equation}
\end{linenomath*}
where $P(R_\mathrm{m}(\theta),t)$ represents the cumulative distribution function calculated under GF conditions
at the moment $t$, i.e.
\begin{linenomath*}
	\begin{equation}
		\label{def:cdf-dyn}
    P(R_\mathrm{m}(\theta)) 
        = \sum_{\mathcal{R}\mathrm{m}
        = \mathrm{R_\mathrm{Lmin}}}^{\mathcal{R}m 
        = \mathrm{R_\mathrm{Umax}}}\wp(\mathcal{R}m(\theta), 
        t)\,.
	\end{equation}
\end{linenomath*}

At this point, we shall introduce three types of rigidity cut off
labeled by three different indexes i.e, $R_\mathrm{C} \rightarrow
\mathrm{R}_{\mathrm{C}(i)}\, $:
\begin{itemize}
	\item $i=0$: for the standard definition of rigidity cutoff,
    i.e. equation (\ref{def:basic-rc}).
	
    \item $i=1$: for rigidity cutoff --secular conditions, i.e. 
    typical time scales greater than one year-- as a function 
    of the cumulative distribution function, i.e. equation
    (\ref{def:cdf}) and (\ref{def:rc-sec-cdf}).
    
	\item $i=2$: for rigidity cutoff --dynamics conditions-- as a 
    function of the cumulative distribution function for the 
    UTC time-stamp, i.e. equation (\ref{def:rc-dyn}) and
    (\ref{def:cdf-dyn}).
\end{itemize}
With these two new types of rigidity cut-off, we shall
redefine, in the next sections, several physical parameters
associated with $\mathrm{R}_{\mathrm{C}(i)}$.

\begin{figure}[!ht]
	\centering
  \includegraphics[trim = 70mm 25mm 70mm 10mm, clip, angle=-90, width=35pc]{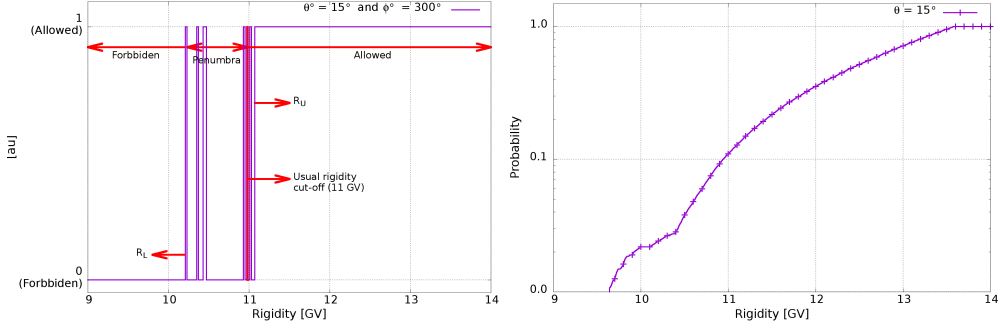}
	\caption{These two plots show the representation of the
  penumbra region. For both, we have calculated --according with
  the method presented in section \ref{subsubsec:rm-calcu}-- the
  magnetic rigidity for a cosmic ray arriving at Bucaramanga,
  Colombia with a zenith angle of $15^\circ$, azimuth angle of
  $300^\circ$ and $t$ corresponding to May
  $13^{\mathrm{th}}$, $2005$. In the left plot, we show the
  standard representation of the penumbra region, i.e. equation
  (\ref{def:basic-rc}), where the violet bars represent the
  intervals for allowed rigidities. The right plot displays our
  interpretation of the penumbra region {\bf averaged for all the
  azimuth angles}, i.e., equations (\ref{def:pdf}) and
  (\ref{def:cdf}): the cosmic ray probability of reaching
  Bucaramanga, Colombia with $\theta=15^\circ$, on May
  $13^{\mathrm{th}}$ of 2005, versus magnetic rigidity
  $R_\mathrm{m}$.}
	\label{rigidity-comparison}
\end{figure}

In the Figure \ref{rigidity-comparison}, we show an example of
the refinement for the estimation of the magnetic rigidity at
Bucaramanga-Colombia, on May 13$^\mathrm{th}$ 2005. In the left
plot, we display the results of the standard method to calculate
the rigidity cutoff (equation (\ref{def:basic-rc})). It is clear
that even in those ``not allowed zones'' (between $10.2$\,GeV and
$11.2$\,GeV) there are several trajectories that could contribute
to the flux at the observation point. This could be particularly
important when it is needed to determine the background flux at
high altitude sites, such as the LAGO site of Mount Chacaltaya at
$5250$\,m a.s.l., or even for the determination of the expected
flux of secondaries impacting
aircrafts\,\citep{Pinilla2015,AsoreyEtAl2017A}. In the same
Figure \ref{rigidity-comparison} (right plot), we illustrate our
new method, displaying $P(R_\mathrm{m}(\theta))$ for different
magnetic rigidities, considering a primary with
$\theta=15^\circ$.

With our method in determining the local directional rigidity
cutoff, it is possible to refine the calculation of the flux of particles
at any observation point while taking into account GF
disturbances, either in long time scales (secular conditions) on
during short term transient phenomena.

\subsection{Estimation of the Primary Flux filtered by
$\mathrm{R}_{\mathrm{C}(i)}$}
\label{sub:step1}

The second step in our simulation chain is to estimate GCR flux
arriving at some geographical point ($112$\,km a.s.l.,
$\mathrm{Lat}, \mathrm{Lon}$) in the area $\mathrm{d}S$, during
time $\mathrm{d}t$, in the solid angle --$\theta$ is the
zenith angle-- $\mathrm{d}\Omega=2\pi \sin(\theta)
\mathrm{d}\theta$, within the energy interval $\mathrm{d}E$ and
with minimum allowed primary momentum of
\begin{linenomath*}
	\begin{equation}
		\label{def:fil-rc}
		p_{\mathrm{min}} = \frac{Ze}{c} 
		\mathrm{R}_{\mathrm{C}(i)}\, ,
	\end{equation}
\end{linenomath*}
with $Z$ the atomic number. Equation (\ref{def:fil-rc})
allows us to filter primaries with insufficient $R_\mathrm{m}$ to
arrive at the point ($\mathrm{Lat}, \mathrm{Lon}, \mathrm{Alt}$).

We estimated the GCR flux, $\Phi$,  at an altitude of $112$\,km
a.s.l., in accordance with the Linsley atmospheric
model\,\citep{NOAA1976}; e.g. at this altitude the mass
overburden vanishes\,\citep{Corsika}, and we approximate $\Phi$
by a simple power law of the form:
\begin{linenomath*}
	\begin{equation}
		\label{gcr-spectrum-noindex}
		\Phi(E,Z,A,\Omega)=
		\frac{\mathrm{d}N(E)}{\mathrm{d}S\,
		\mathrm{d}\Omega \, \mathrm{d}t \,
    \mathrm{d}E} \simeq j_0(Z,A) \left(\frac{E}{E_0}\right)^{\alpha(E,Z,A)}\,,
	\end{equation}
\end{linenomath*}
where the spectral index ($\alpha(E,Z,A)$) can be considered
constant with respect the energy, i.e.
$\alpha(E,Z,A)~\approx~\alpha(Z,A)$, from $10^{11}$\,eV to 
$10^{15}$\,eV\,\citep{Letessieretal2011} and $E_0$ has a
value of $10^{12}$\,eV. For each type of GCR considered,
$\alpha(Z,A)$, is individualized by its mass number (A) and
atomic number (Z). Finally, $j_0(Z,A)$ is the normalization
parameter. Both, the spectral indices and $j_0$, have been
obtained from the compilation produced
by\,\citep{Wiebel-Sooth-etal1998}.

We calculated $\Phi$ using the fact that multiple observations
have confirmed that at low energies ($E\lesssim 5.5\times
10^{19}$\,eV) the GCR flux can be considered isotropic\,\citep[see
e.g.][]{Abraham2007} and, in this case, equation
(\ref{gcr-spectrum-noindex}) is integrated to obtain the expected
number of primaries for every nuclei $(Z,A)$ as:
\begin{linenomath*}
	\begin{equation}
		\label{number-by-chem}
		N(Z,A,\theta) = 
		\left ( \pi \Delta S \Delta t
		\sin^2(\theta) \right ) 
		j_0(Z,A) 
    \frac{\left(E/E_0\right)^{\alpha(Z,A) 
		+ 1}}{\alpha(Z,A) + 1}
		\Big|_{E_{\min}}^{E_{\max}}\, ,
	\end{equation} 
\end{linenomath*}
with $E_{\max} - E_{\min} \equiv \Delta E$ as the energy gap,
which, in our case, varies from a few GeV
($E_{\min}$) up to $10^6$\,GeV ($E_{\max}$)\,\citep{Asorey2011}.
It is clear that the first factor depends only on the
zenith angle $\theta$, and so, $\mathcal{N}(\theta) \equiv
\sin^2(\theta) \pi \Delta S \Delta t$. Thus, equation
(\ref{number-by-chem}) can be expressed as
\begin{linenomath*}
	\begin{equation}
		\label{number-by-chem-0}
		N(Z,A,\theta) = \mathcal{N}(\theta) j_0(Z,A)
    \frac{\left(E/E_0\right)^{\alpha^\prime(Z,A)}}{\alpha^\prime(Z,A)}
		\Big|_{E_{\min}}^{E_{\max}}\, ,
	\end{equation}
\end{linenomath*}
where $\alpha^\prime(Z,A)=\alpha(Z,A) + 1$. For the calculation
of $N(Z,A,\theta)$ we have used: $\Delta S~=~1$\,m$^2$, $\Delta
t=86400$\,s, i.e., at least one day of the primary flux per square
meter for each primary in the range $1\leq Z \leq 26$, for
$\theta$ from $0^\circ$ to $90^\circ$. The $N(Z,A,\theta)$ will
then be filtered via $E_\mathrm{\min}$ according to equation
(\ref{def:fil-rc}).

This means that we can identify three different kinds of primary fluxes,
one per each different GF conditions ${R_\mathrm{C}(i)}$: 
\begin{itemize}
  \item $\Phi_{(0)}$ for $R_{\mathrm{C}(0)}$, i.e.
    $p_\mathrm{\min} c = Ze \times R_{\mathrm{C}(0)}$.
        
  \item $\Phi_{(1)}$ for $R_{\mathrm{C}(1)}$, i.e.
    $p_\mathrm{\min} c = Ze \times R_{\mathrm{C}(1)}$.
        
  \item $\Phi_{(2)}$ for $R_{\mathrm{C}(2)}$, i.e.
    $p_\mathrm{\min} c = Ze \times R_{\mathrm{C}(2)}$.
\end{itemize}
Thus, the number of particles given by 
(\ref{number-by-chem-0}) will be susceptible to $R_\mathrm{m}$
corrections by the modification of the local rigidity cutoff, and
it can be re-written as
\begin{linenomath*}
	\begin{equation}
		\label{number-by-chem-index}
		N_{(i)} = \mathcal{N}(\theta) j_0(Z,A)
    \frac{\left(E/E_0\right)^{\alpha^\prime(Z,A)}}{\alpha^\prime(Z,A)}
	\end{equation}
\end{linenomath*}
Here, the subindex $i$ of any quantity denotes the type of effect included.

As value of $E_{\max}$ we have used $10^6$\,GeV, because at these energies 
the flux is so low that it can not affect the
distribution of the secondary background at ground level. It is
important to stress that, for a given point, $E_{\min(i)}$
depends on the primary $Z$, theta arrival direction $(\theta)$
and time, i.e., $E_{\min(i)} \equiv
E_{\min(i)}(Z,\theta,t)$. 

When GF corrections are calculated with our method, the new
expected primary flux $\Phi_{(1)}$ obtained from equation
(\ref{number-by-chem-index}) will depend on the arrival direction
of each primary\,\citep{Dorman-etal2008}.

\subsection{Estimation of flux of secondary particles at 
ground level corrected by the Geomagnetic Field}
\label{sub:step2}

The following step of the simulation chain is the correction for
the effect of the Geomagnetic Field on the flux of secondary
particles at ground level. As was mentioned in section
\ref{LAGO}, one of the main objectives of this simulation chain
is to calculate the expected flux of secondaries at the detector
level at any site of the LAGO network. Once the primary flux
$\Phi_{(i)}$ is calculated, the second step is to determine the
interactions of those primaries with the atmosphere. This
simulation step is performed with the CORSIKA code\,\citep{Corsika} (Currently,
CORSIKA v7.3500, compiled using the following options:
QGSJET-II-04\,\citep{Ostapchenko2011}; GHEISHA-2002; EGS4; curved
and external atmosphere, and volumetric
detector). The local geomagnetic field values
$B_x$ and $B_z$ needed by CORSIKA to account for GF effects
during particles propagation in the atmosphere are obtained from
the IGRF-11 model. Secondary particles are tracked to the lowest energy threshold
allowed by CORSIKA for each type (currently, $E_s\geq 5\times
10^{-2}$\,GeV for $\mu^\pm$ and hadrons (excluding $\pi^0$), and
$E_s \geq 5\times10^{-5}$\,GeV for $e^\pm, \pi^0$ and $\gamma$)
to get the most comprehensive distribution of secondaries at each
site. In this work, the atmosphere at each site was
simulated by using profiles of the applicable 
MODTRAN atmospheric model\,\citep{Kneizys1996} provided with
CORSIKA. For the Bucaramanga site we use a tropical profile and for
San Carlos de Bariloche a midlatitude summer profile. Currently, the LAGO
collaboration is developing and validating a method to obtain the
local atmospheric profiles for each LAGO site during different
weather conditions based on the Global Data Assimilation System
(GDAS)\,\citep{NOAA2009}, as differences have been observed 
between generic MODTRAN models and balloon measurements at the
planed LAGO site in Antarctica\,\citep{DassoEtal2015}.

A large number of primary showers need to be simulated (typical
values are of several billions of showers for 24\,h of flux at
each site). A set of local clusters have been deployed and tuned
for this particular calculation. These are maintained at some
institutions of the LAGO Collaboration. This simulation chain has
been also integrated into a dedicated Virtual Organization,
{\textit{lagoproject}}, as part of the European Grid
Infrastructure (EGI, http://www.egi.eu) activities. The Grid
implementation of CORSIKA was deployed with two ``flavors'',
which run by using GridWay Metascheduler
(http://www.gridway.org/doku.php)\,\citep{HuedoMonteroLlorente2001}
or with a second approach through a Catania Science Gateway
interface\,\citep{BarberaFargettaRotondo2011}. Massive
calculations can be executed with the former, via the
Montera\,\citep{RodriguezMayoLlorente2013}, the
GWpilot\,\citep{RubioEtal2015}\, or the 
GWcloud\,\citep{RubioHuedoMayo2015} frameworks.

In the Science Gateway approach a user can seamlessly run a code
on different infrastructures by accessing a unique web-based
entry point with an identity provision. Users only have to upload
the input data or invoke a PID (persistent identifier or
reference to a digital set of files) and click on the run icon.
The final result will be retrieved whenever the job has ended.
The underlying infrastructure is absolutely transparent to the
user and the system decides on which sites and computing platform
the code will be compiled and run\,\citep{RodriguezEtal2015,
AsoreyEtal2016}.

To deal with the computational complexity introduced by the
refinement described in the previous subsection, we built a
library for each site containing the simulated particles starting
from a very low momentum primary threshold of $\sim (350 \times
Z)$\,MeV$/c$ (i.e. $1$\,GeV of total energy for protons). Each
secondary impinging the detector is tagged with information from
its parent primary particle, which allows the calculation of its
magnetic rigidity $R_\mathrm{m}$. Then, because each
secondary at the ground comes from some primary impinging at the
atmosphere, from the $R_{\mathrm{C}(i)}$ obtained for each
condition according to equations (\ref{def:fil-rc}) and
(\ref{number-by-chem-index}), we are able to determine if each
secondary would reach the detector under that particular GF
condition.

\section{Results for the LAGO sites of Bucaramanga, Colombia 
and San Carlos de Bariloche, Argentina}
\label{results}

\subsection{Magnetic Rigidity}

\begin{figure}[h]
	\centering
  \includegraphics[trim = 70mm 25mm 70mm 10mm, clip, angle=-90, width=35pc]{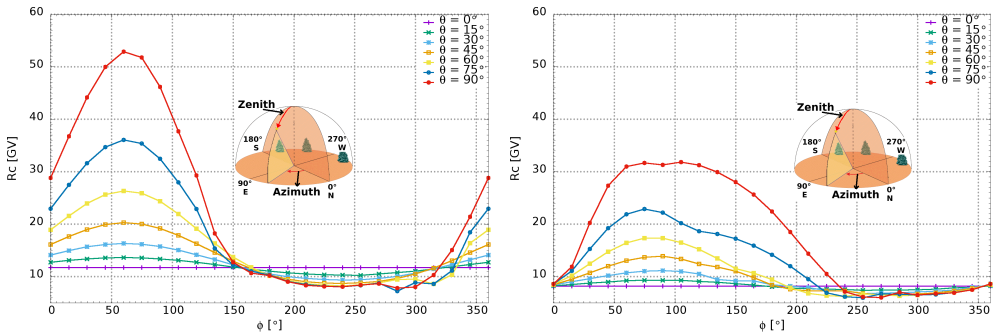}
	\caption{
    Calculation of $R_{\mathrm{C}(0)}$ (GF secular conditions,
    equation (\ref{def:basic-rc})) at the atmosphere edge
    ($112$\,km a.s.l.), as a function of the azimuth angle $\phi$
    and for different zenith angles $\theta$. This calculation is
    presented for two LAGO sites: Bucaramanga, Colombia, left,
    and San Carlos de Bariloche, Argentina, right. 
    The strong dependence between $R_{\mathrm{C}(0)}$  and the arrival directions 
    at both cities, is evident.
		}
	\label{figRcSecBGABRC}
\end{figure}

As we mentioned before, we applied our simulation chain to the
location of two LAGO sites: Bucaramanga, Colombia (BGA) and San
Carlos  de Bariloche, Argentina (BRC). Results for the standard
rigidity cutoff calculation, $R_{\mathrm{C}(0)}$, are displayed
in Figure \ref{figRcSecBGABRC} for each site. As expected, there
is a strong dependence between $R_{\mathrm{C}(0)}$ and the
arrival directions at both cities, which induces a noticeable
decrease in the number of GCRs producing secondary particles at
ground level. For Bucaramanga, it is interesting to mention the
oddity of the behavior of the rigidity cutoff for large azimuthal
($250^{\circ} \lesssim \phi\lesssim 300^{\circ}$) and zenithal
($45^{\circ} \lesssim \theta \lesssim 90^{\circ}$) angles. We
have backtracked several incoming trajectories and discovered
that this anomaly in the rigidity cutoff seems to be associated
with the deflection of particles with low $R_\mathrm{m}$, whose
trajectories cross zones with high gradients of the
GF\,\citep{Suarez2015a}.

\begin{figure}[h]
	\centering
	\includegraphics[trim = 70mm 25mm 70mm 10mm, clip, angle=-90, width=35pc]{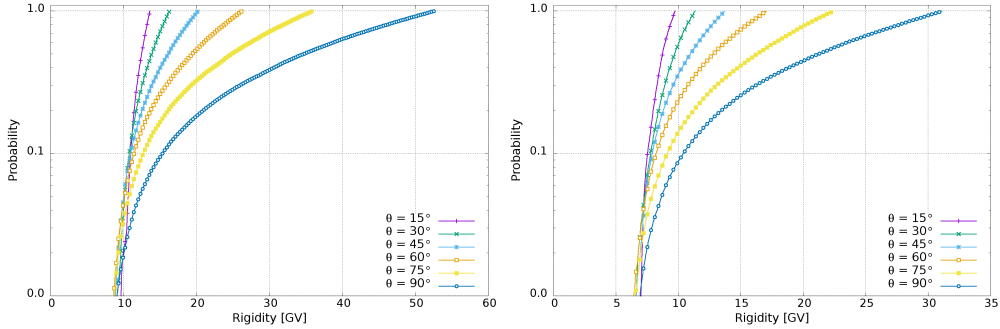}
	\caption{Magnetic rigidity calculated on the penumbra region as
  a cumulative distribution function, according to equation
  (\ref{def:cdf}) for different arriving zenith angles at the
  sites of Bucaramanga (left) and San Carlos de Bariloche
  (right).}
	\label{asymptotic-rc-stat}  
\end{figure}

The cumulative probability distributions (\ref{def:cdf}), as
functions of the magnetic rigidity for various zenith angles, are
displayed in Figure \ref{asymptotic-rc-stat}. There, lower
magnetic rigidities are associated with particle trajectories
having small zenith angles. Notice that both plots are
qualitatively different and this probably is evidence of the
complexity of the GF present at the two very distinct latitudes.

\subsection{Primary Flux Corrected by GF}

\begin{figure}[h]
	\centering
	\includegraphics[trim = 70mm 25mm 70mm 10mm, clip, angle=-90, width=35pc]{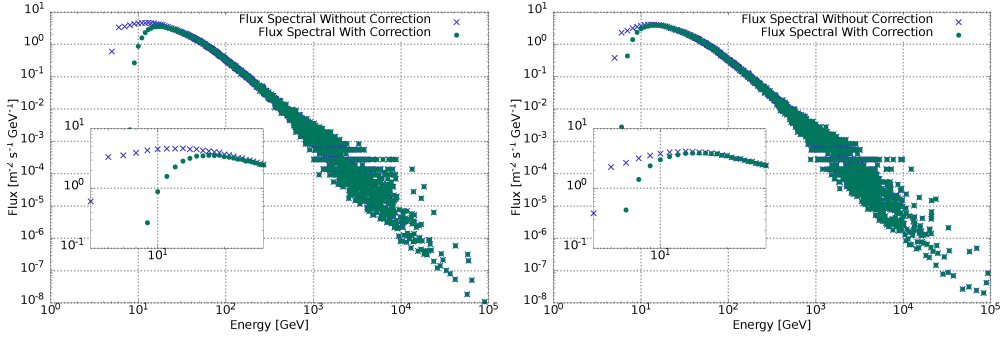}
	\caption{
		Expected GCRs flux at 112 km a.s.l. producing secondaries at
    ground level with only the standard corrections
    ($\Phi_{(0)}$, blue crosses) and with the proposed method
    in secular conditions ($\Phi_{(1)}$, green squares), for
    Bucaramanga (left) and San Carlos de Bariloche (right). The
    influence of the GF corrections is evident at the low energy
    limits.}
    \label{en-sec-enerpri-con} 
\end{figure}

As was explained in section \ref{subsubsec:inter-penum}, once the
penumbral CDFs is obtained, it is possible to refine the
calculation of the expected primary flux $\Phi_{(i)}$ and the
corresponding flux of background secondaries $\Xi_{(i)}$ at
ground level. In Figure \ref{en-sec-enerpri-con} the GCR flux
$\Phi_{(0)}$ and $\Phi_{(1)}$ are displayed for both LAGO sites:
BGA and BRC. Only those primaries producing secondaries at ground
level are shown. In both cases, the influence of the GF
corrections is only significant at lower energies, $\mathbf{\sim
15}$\,GeV. As expected, instead of a sharp cutoff as in the
standard case, a smooth cutoff is observed, corresponding to the 
different rigidities cutoff in the different regimes, and the
flux of primaries is affected according with figure
\ref{asymptotic-rc-stat}, i.e., a bigger effect for BGA (rigidity
up to $\sim50$\,GV) than BRC (rigidity up to $\sim30$\,GV).

The primary flux interacts with the atmosphere and produces the
secondary flux $\Xi_{(i)}$ at the ground level. These
interactions were simulated by CORSIKA obtaining a very
comprehensive distribution of particles at the detector level. To
estimate the response of the WCD to each type of secondary
particle, this flux is analyzed using a detailed Geant4
simulation of the LAGO detector will be described and the first
preliminary results were showed by \citet{Otiniano2015}.

\subsection{Secondary Flux Corrected by GF}

\begin{figure}[h]
	\centering
	\includegraphics[trim = 70mm 25mm 70mm 10mm, clip, angle=-90, width=35pc]{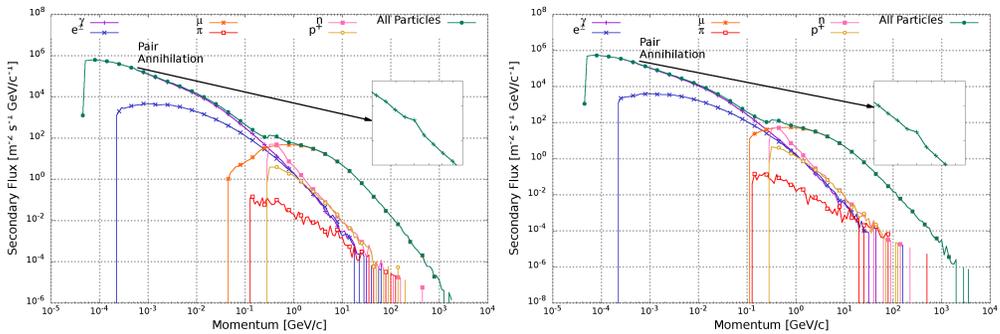}
	\caption{The expected secondary spectra at detector level
  $\Xi_{(1)}$ is shown for Bucaramanga (BGA, left) and San Carlos
  de Bariloche (BRC, right), for different type of secondaries:
  photons (purple), electrons and positrons (blue), muons
  (orange), pions (red), neutrons (pink), protons (gold), total
  spectrum of secondaries (green).}
	\label{secondary-flux} 
\end{figure}

Figure \ref{secondary-flux} displays the simulated spectra of
secondaries $\Xi_{(1)}$ (under secular conditions) at both
cities. A noticeable peak for the distribution of secondary
neutrons and protons are evident at both sites. At these low
altitudes, a muon hump is also visible in the distribution
spectra, and this is typically used as a calibration point for
WCD\,\citep[see e.g.][]{AEtchegoyen-etal2005, Asorey2013}.


\begin{figure}[h]
  \centering
  \includegraphics[trim = 70mm 25mm 70mm 10mm, clip, angle=-90, width=35pc]{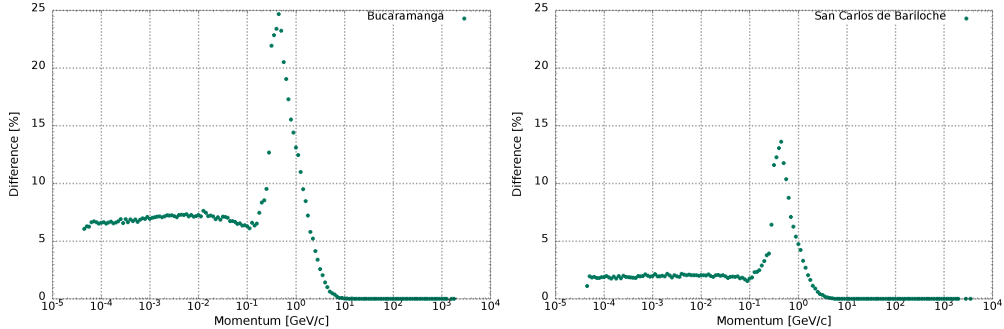}
  \caption{Percentage Flux Difference of secondaries at the
  ground between the flux calculated for for the standard
  definition of rigidity cutoff (equation (3)) and for rigidity
  cutoff secular conditions (equation (8)), i.e. $\Delta
  \Xi_{0-1}$: BGA (left) and BRC (right). It clearly peaks at
  $\sim 500$\,MeV/c and is framed between $100$\,MeV/c and
  $3$\,GeV.}
	\label{en-sec-espec-rigi-ground} 
\end{figure}

By defining the \textit{flux percentage difference}, $\Delta
\Xi_{i-j}$, it is possible to get a better understanding of the
energy range where the geomagnetic corrections are more
important, specially when dynamic variations are considered.
Thus
\begin{linenomath*}
	\begin{equation}
		\label{FluxDiference}
		\mathrm{Difference}_{\%} \equiv \Delta \Xi_{i-j} 
		= 100 \left ( \frac{\Xi_{(i)} -\Xi_{(j)}}{\Xi_{(i)}} 
		\right ) \%\, ,
	\end{equation}
\end{linenomath*}
where $i,j$ are the indices corresponding to the configuration
of the GF introduced in the section \ref{subsubsec:inter-penum}.
To evaluate the impact of this new method, the differences
between cases $(0)$ (standard calculation) and $(1)$ (new
method), as a function of the secondary particles momentum, are
illustrated in Figure \ref{en-sec-espec-rigi-ground}. The
presence of a peak at $\sim500$\,MeV/c is evident for both sites
in the $\Delta \Xi_{0-1}$ distribution, located between
$100$\,MeV/c and $\sim 3$\,GeV/c. For lower energies, the
difference is a bit larger at BGA than at BRC, as we expected
after figure \ref{figRcSecBGABRC}. When we explored in more
detail the particle component of the secondaries at these
momenta, we found that these differences are dominated by
secondary neutrons\,\citep{Suarez2015a}, where the diminution is
of the order of $35\%$. This result, in our simulation, agree
with the fact that variations in the flux registered by Neutron
Monitors are a proxy of the changing conditions in the near-earth
space environment. For energies higher than $\sim10$\,GeV,
corrections are not important.

\begin{figure}[htb!]
  \centering
  \includegraphics[trim = 20mm 55mm 20mm 55mm, clip, width=35pc]{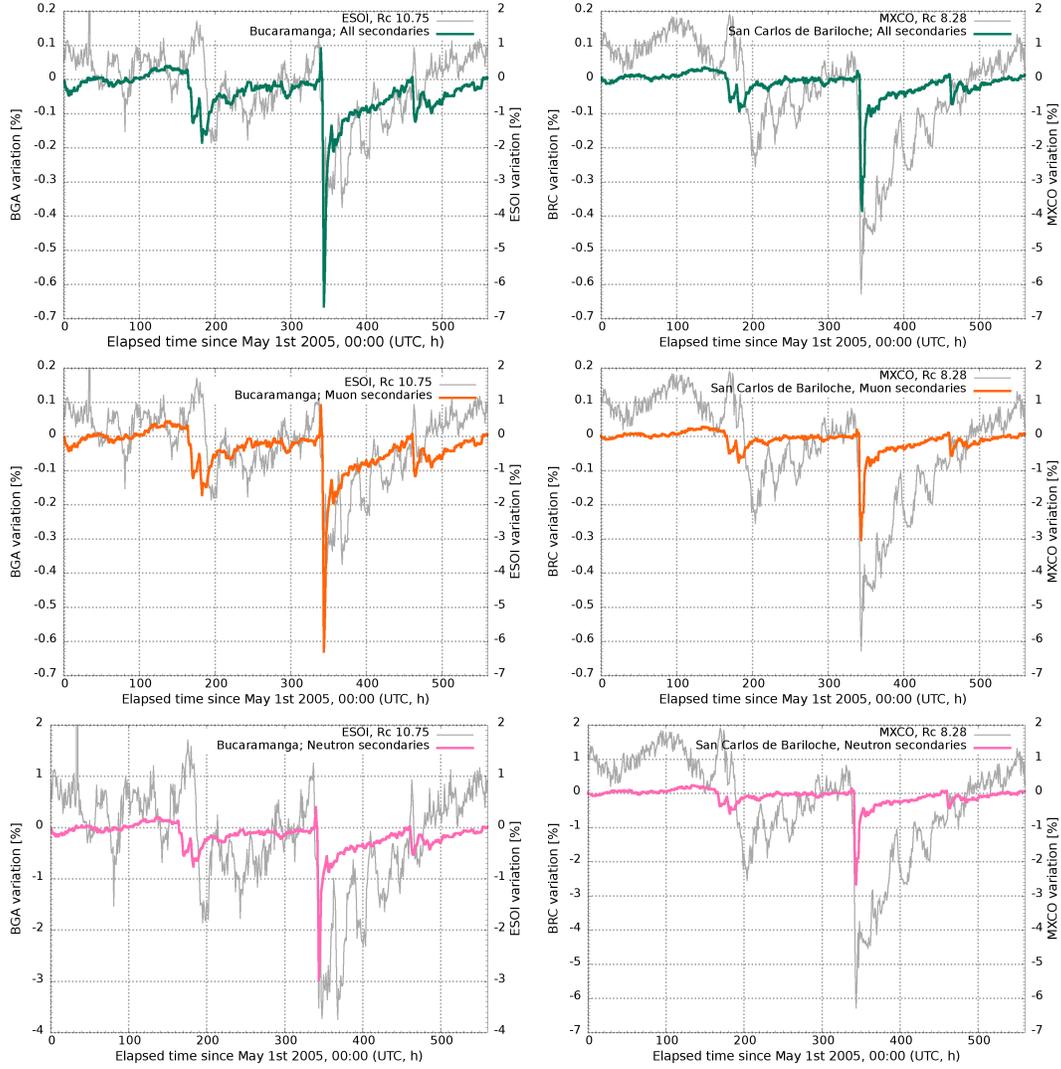}
	\caption{Time evolution of the expected flux of secondaries
  during dynamic conditions of the GF, $\Delta \Xi_{1-2}$ for May
  of 2005 at Bucaramanga (left) and at San Carlos de Bariloche
  (right). On background, in gray, is presented the neutron flux
  is shown, 
  registered by two Neutron monitors with similar
  $R_{\mathrm{C}(0)}$: to BGA, ESOI with
  $R_{\mathrm{C}(0)}=10.75$\,GV (left); to BRC, Mexico with
  $R_{\mathrm{C}(0)}=8.28$\,GV (right). The data for the neutron
  monitor was taken from http://www.nmdb.eu. The first row
  corresponds to the total flux of particles at ground level,
  while in the second row we illustrate the evolution of the muon
  flux $\Delta \Xi_{1-2}^\mu$. The third one displays the neutron
  flux $\Delta \Xi_{1-2}^n$. Note the scale difference on the y-axis
  for each plot.}
	\label{en-sec-filt-dst-all-a}
\end{figure}

Finally, this tool allows the study of the impact of dynamic
conditions of the GF in the distribution of secondary particles
by comparing secular conditions of the GF with the evolution of
the GF states as a function of time, $\Delta \Xi_{1-2}$. This is
because the calculations performed by this method are focusing on
the background of the GCR flux, e.g., we did not consider the solar
particle event during the geomagnetic storm of May 13-17, 2005,
but just the influence of the state of the GF, for this UTC-time,
over the background of GCR flux. This is shown in
Figure \ref{en-sec-filt-dst-all-a}, where the time evolution of
$\Delta \Xi_{1-2}$ is displayed at both sites for May of 2005. We
have selected this particular month because the strong
geomagnetic storm on May 13-17, 2005, generated intense
disturbances in the GF\,\citep[see e.g.][] {Adekoya2012,
Bisi-etal2010, Galav-etal2014}. Three
particular cases are shown: the total flux secondary particles,
$\Delta \Xi_{1-2}$, the muon flux $\Delta \Xi_{1-2}^\mu$ and the
neutron flux $\Delta \Xi_{1-2}^n$. It is clear that, beside the
time coincidence of the flux variations, it is more significant
at Bucaramanga than at Bariloche, and that the neutron flux at 
ground level is the most affected component by the GF
activity, which reinforces the known sensitivity of this
particular constituent for the observation of geomagnetic
disturbances\,\citep[see e.g.][]{Belov2005}.

As reference, on the background of each sub-plot in figure
\ref{en-sec-filt-dst-all-a} (gray line) the flux of neutrons at
ground level is shown, registered by two Neutron Monitors (NM)
with similar rigidities to both sites;  i.e. $R_{\mathrm{C}(0)}$
of $10.75$\,GV for NM of ESOI and $8.28$\,GV for NM of Mexico. As
reference, the $R_{\mathrm{C}(0)}$ for BGA is $11$\,GV and for
BRC is $8.1$\,GV. Both NM show a decrease between $300$ and $400$
in elapsed UTC time, that is in coincidence with our simulation
results. Because we have simulated the effect of the GF under GCR
flux, i.e., we do not simulate solar particle events, it is
possible, with our approach, to estimate the contribution of the
GF topology to a Forbush decrease event.
\newpage

\newpage


\section{Final remarks}
\label{conclusions}

In this paper, we have presented the LAGO space weather chain of
simulations devised to obtain precise calculations of secondary
particle flux at ground level that can be used at every
geographical position. It takes into account geomagnetic
corrections for both secular (long term phenomena with typical
time scale larger than a year) and transient events
(with typical time scales of hours to days). We shall consider
all calculations performed without this new method as a first
approximation to the more precise determination of the real flux
of secondaries particles, calculated when the effects of the
geomagnetic field are fully considered.

Variations of the flux of secondaries at two LAGO sites with
different latitude/longitude (Bucaramanga, $7^\circ\,8^{'}$N,
$73^\circ\, 0^{'}$W and San Carlos de Bariloche
$41^\circ\,9^{'}$N, $71^\circ\, 18^{'}$W) were calculated for
both secular geomagnetic conditions and under transient events,
during the geomagnetic active month of May 2005. Our simulations
show that the secondary flux is sensitive to the latitude and
that the secondary neutrons at the ground level are the most
affected flux component due to variations of the geomagnetic
field during space weather phenomena. While our calculation
relies on the isotropy of the GCR flux, it is important to note
that during certain FDs, small anisotropies in the flux of
primaries could be induced by the configuration of the incoming
magnetic cloud and the disturbances of the geomagnetic field
during these particular events. Actually a $\sim 1\%$ anisotropy
in the flux of secondary muons was observed at ground level
during the Forubush decrease of December 13,
2006\,\citep{Kane2006, Fushishita-etal2010}. However, since our
WCD are not sensitive to the arrival direction of secondary
particles, we will not be able to detect such small effect while
the total flux of secondaries remains constant.

Several dedicated clusters and a Grid-based implementation have
been deployed for these calculations. A dedicated Virtual
Organization, {\textit{lagoproject}}, part of the European Grid
Infrastructure (EGI, http://www.egi.eu) activities has been
created, and available tools for Grid have been adapted and 
implemented to run CORSIKA in a absolutely transparent way to 
the user.

The standard definition of the penumbra region for magnetic
rigidities generates a complex structure of particle
trajectories: permitted, prohibited and quasi-trapped orbits,
which does not allow to derive all values for the
$R_\mathrm{m}$\,\citep{SmartEtal2006}. Currently, calculations of
rigidity cutoff tend not to consider the effects involved in the
penumbra, and always use a single effective value
(equation (\ref{def:basic-rc})) to account and characterize
all the complexity of involved effects\,\citep{SmartShea2009}. In
this paper the concept of rigidity cutoff $R_\mathrm{C}$ has been
generalized as a time dependent function of the cumulative
probability distribution (see equation (\ref{def:cdf-dyn})). With
this refinement, at the penumbra region, we can obtain a
non-vanishing probability to have an incoming particle (with a
zenithal angle $\theta$) contributing to the flux of primaries at
the observation point.

Combining the data measured at different locations of the LAGO
detection network, with those obtained from the detailed
simulation performed by this space weather chain, we are now
capable of providing a better understanding of the temporal
evolution and of the small and large scales disturbances of the
space weather phenomena.

\acknowledgments
The authors thanks the enlightening suggestions and criticism
from the anonymous referees and also from the AGU Space Weather
Editorial Office which have helped to make this work clearer and
more focused.  We appreciate the support of Vicerrector\'{\i}a
Investigaci\'on y Extensi\'on Universidad Industrial de Santander
for its permanent sponsorship, and acknowledge the financial
support of Departamento Administrativo de Ciencia, Tecnolog\'ia e
Innovaci\'on of Colombia (COLCIENCIAS) under contracts
FP44842-051-2015 and FP44842-661-2015. The authors acknowledge
the support of COLCIENCIAS, CONICET and MINCyT for funding
bilateral cooperation Argentina-Colombia, grant AR:CO-15/02
CO:729-2015. HA and MSD acknowledge the support from Inn\'ovate
Per\'u, grant 398-PNICP-PIBA-2014.  Significant parts of the
calculations needed for this work was performed with the
computational support of the Centros de Supercomputaci\'on y
C\'alculo Cient\'ifico de la Universidad Industrial de Santander.
We also acknowledge the NMDB
database (www.nmdb.eu), founded under the European Union's FP7
programme (contract no. 213007) for providing the data. Neutron
monitor of the Emilio Segre Observatory is supported by
collaboration ICRC-ESO (Tel Aviv University and Israel Space
Agency, Israel) and University ``Roma Tre'' with IFSI-CNR
(Italy). Neutron monitor data of Mexico City is provided by the
Cosmic Ray Group of the Geophysical Institute at the Universidad
Aut\'onoma de M\'exico (UNAM). The authors are grateful to the
LAGO and Pierre Auger Observatory Collaboration
members (http://lagoproject.org/collab.html) for their
continuous engagement and support.
\newpage


\end{document}